\documentclass[12pt,preprint]{aastex}

\newcommand\etal{\hbox{et~al.}}

\newcommand{\Lbol}{\mbox{$L_{Bol}$}}
\newcommand\Rstar{\hbox{$R_*$}}
\newcommand\Teff{\hbox{$T_{eff}$}}
\newcommand\logG{\hbox{log $ g$}}
\newcommand\Vesc{\hbox{$V_{esc}$}}
\newcommand\Rsun{\hbox{$R_\odot$}}

\newcommand\vinf{\hbox{$v_\infty$}}
\newcommand\kms{\hbox{$km ~ s^{-1}$}}
\newcommand\Mdot{{\hbox{$\dot M$}}}

\newcommand\TX{\hbox{$T_{\rm X}$}}
\newcommand\FX{\hbox{$F_{\rm X}$}}
\newcommand\JX{\hbox{$J_{\rm X}$}}
\newcommand\LX{\hbox{$L_{\rm X}$}}
\newcommand\LXLB{\hbox{$L_{\rm X}/ L_{\rm {Bol}}$}}

\newcommand{\Halpha}{\mbox{H$\alpha$}}
\newcommand\HeI{\hbox{He {\sc i}}}
\newcommand\HeII{\hbox{He {\sc ii}}}
\newcommand\NVII{\hbox{N {\sc vii}}}
\newcommand\OVII{\hbox{O {\sc vii}}}
\newcommand\MgXI{\hbox{Mg {\sc xi}}}
\newcommand\NeIX{\hbox{Ne {\sc ix}}}
\newcommand\SiXIII{\hbox{Si {\sc xiii}}}
\newcommand\SXV{\hbox{S {\sc xv}}}

\newcommand\CIV{\hbox{C {\sc iv}}}
\newcommand\CV{\hbox{C {\sc v}}}

\newcommand\PV{\hbox{P {\sc v}}}
\newcommand\PVI{\hbox{P {\sc vi}}}

\newcommand\SV{\hbox{S {\sc v}}}
\newcommand\SVI{\hbox{S {\sc vi}}}

\newcommand\DEMC{\hbox{$\Delta EM_C$}}
\newcommand\FXFS{\hbox{$F_X / F_*$}}
\newcommand\QPV{\hbox{$q$ (P {\sc v})}}
\newcommand\QSV{\hbox{$q$ (S {\sc v})}}

\shorttitle{XUV and the PV Problem}
\shortauthors{Waldron \& Cassinelli.}

\begin{document}
\title{The Importance of XUV Radiation as a Solution to the \PV\ Mass Loss Rate Discrepancy in
O-Stars}
\author {W.~L. Waldron$^{1}$ and J.~P. Cassinelli$^{2}$\\
$^{1}$Eureka Scientific Inc., 2452 Delmer St., Oakland CA, 94602;
wwaldron@satx.rr.com\\
$^{2}$Dept. of Astronomy, University of Wisconsin-Madison, Madison, WI 53711;
cassinelli@astro.wisc.edu}

\begin{abstract}
A controversy has developed regarding the stellar wind mass loss rates in O-stars. 
The current consensus is that these winds may be clumped which implies 
that all previously derived mass loss rates using density-squared diagnostics 
are overestimated by a factor of $\approx$ 2. 
However, arguments based on {\it FUSE} observations of the \PV\ resonance line doublet
suggest that these rates should be smaller by another order of magnitude, 
provided that \PV\ is the dominant phosphorous ion among these stars. 
Although a large mass loss rate reduction would have a range of undesirable consequences,
it does provide a straightforward explanation of the unexpected symmetric and un-shifted X-ray
emission line profiles observed in high energy resolution spectra. 
But acceptance of such a large reduction then leads to a contradiction with an
important observed X-ray property: the correlation between He-like ion source radii and their
equivalent X-ray continuum optical depth unity radii. Here we examine the phosphorous
ionization balance since the \PV\ fractional abundance, \QPV, is fundamental to understanding
the magnitude of this mass loss reduction. We find that strong ``XUV'' emission lines in the
\HeII\ Lyman continuum can significantly reduce \QPV. Furthermore, owing to the unique energy
distribution of these XUV lines, there is a negligible impact on the \SV\ fractional abundance (a
key component in the {\it FUSE} mass loss argument). We conclude that large reductions in 
O-star mass loss rates are not required, and the X-ray optical depth unity relation remains valid.
\end{abstract}

\section{Introduction}
Over the past several years, questions concerning the validity of what we refer to as the
``traditional'' O-star mass loss rates, \Mdot, have arisen owing to the predictions of clumped
wind models (see recent Potsdam Workshop, Hamann \etal\ 2008, and references therein).
Abbott \etal\ (1981) showed that inhomogeneous winds lead to an enhancement in density-
squared emission processes such as \Halpha, infrared and radio free-free, which means that the
inferred \Mdot\ would be overestimated (see also Puls \etal\ 2006). Although a clear picture of
these clumpy structures is still being developed, the so-called ``clumping factor'' ($f_{cl}$) is
believed to be $\approx$ 4 to 5, and the reduction in \Mdot\ scales with $\sqrt f_{cl}$ (see
General Discussion in Hamann \etal\ 2008). 

The derived \Mdot\ from diagnostics that are linearly dependent on density (e.g., analyses of
unsaturated UV resonance line profiles) are expected to be independent of clumping effects (Puls
\etal\ 2006) and supposedly should provide more reliable \Mdot\ values. In fact, analyses of the 
\PV\ resonance line doublet ($\lambda \lambda$ 1118, 1128 \AA) obtained from {\it FUSE}
observations led to the conclusion that traditional \Mdot\ are overestimated  by a factor of 10 or
more (Massa \etal\ 2003, hereafter M03; Fullerton \etal\ 2006; hereafter F06). Such reductions
have far-reaching consequences.  For example, Hirschi (2008) concluded that the well known
evolutionary tracks of massive stars could survive \Mdot\ reductions by a factor of 2, but not by a
factor of 10 or more.  

Resolution of this \Mdot\ problem is also of particular importance with regards to the observed
X-ray emission line properties obtained from $Chandra$ and $XMM-Newton$ observations. 
Waldron \& Cassinelli (2007) analyzed the {\it Chandra HETGS} X-ray line properties for a large
number of OB stars, and two of their conclusions are directly relevant to this \Mdot\ issue. 
1) All of the resolved X-ray emission lines are very broad (i.e., HWHM range of 300 to 1000
\kms), symmetric, and the majority have minimal line-shifts. 
Although Waldron \& Cassinelli (2001) were the first to demonstrate that these X-ray line profiles
could in fact be easily explained by a significant reduction in \Mdot, this was inconsistent with the
then accepted observed \Mdot, and they suggested a clumpy or non-symmetric wind structure as
a possible explanation. 
2) The radial locations of the He-like $fir$ (forbidden, intercombination, resonance) line sources,
as derived from their $f/i$ line ratios, typically range from 1.2 to 10 \Rstar,
and these distances are well correlated with their respective stellar wind X-ray continuum
optical depth unity radii which we shall refer to as the ``X-ray continuum optical depth unity 
relation'' (hereafter abbreviated as XODUR). This relation implies that the traditional \Mdot\ (like
those of Vink \etal\ 2000) must be correct (i.e., within a factor of 2 or so). Nevertheless, large
\Mdot\ reductions have become a widely used explanation of the X-ray line profile symmetry
problem (e.g., Kramer \etal\ 2003; Cohen \etal\ 2006) because the emission from both the near
and far side of a star would be observable in an optically thin wind. The subject is not yet resolved
since Oskinova \etal\ (2004, 2006) finds that the symmetry can also be explained by accounting
for the porosity of clumped (or fragmented) winds without the need for a reduction in \Mdot, but
Owocki \& Cohen (2006) present arguments against the porosity influence. 

The primary goal of this paper is to examine the effects of an excess of hard radiation within the
\HeII\ Lyman continuum on the ionization equilibrium of phosphorus by utilizing {\it only} outer
shell photoionization processes. This is important since the
proposed large reduction in \Mdot\ is based entirely on the assumed \PV\ fractional ionization
abundance. In addition, we also consider the sulfur ionization balance because of two arguments
used by M03 and F06: 1) since phosphorous and sulfur have overlapping ranges in ionization
energy, the dominant stages of sulfur can be used as surrogates for the corresponding ones of
phosphorous, and; 2) both \PV\ and \SV\ are likely to be the dominant ionization stages
throughout the O-star spectral range. 

We propose that the primary source of this excess hard radiation is produced by the ``XUV''
spectral energy band which has long been known to have the capability to produce anomalously
higher wind ionization stages (e.g., Waldron 1984; MacFarlane \etal\ 1994; Pauldrach \etal\ 1994,
2001). Based on solar physics studies, the XUV lower energy bound seems to be rather loosely
defined, but the upper limit appears to be fixed at 124 eV (100 \AA) {\footnote{Withbroe \&
Raymond (1984) define a XUV energy range from 25 eV (\HeI\ edge) to 124 eV.}} where the
XUV upper limit represents the start of the X-ray energy band. For our purposes, we adopt an
XUV radiation energy band defined as 54.4 eV (\HeII\ edge) to 124 eV, since below the \HeII\
edge, the radiation is dominated by photospheric emission. Although XUV radiation is 
un-detectable in O-stars, its effects can be manifested by studying other spectral bands.  In
particular, with regards to the two key arguments used by M03 and F06, we show that XUV
radiation produces dissimilar changes in the phosphorous and sulfur ionization equilibria.

\section{Importance of XUV Radiation}
In this section we use graphical arguments to illustrate the importance of XUV radiation with
respect to the total  radiation cooling curve and the ionization equilibria of phosphorous and
sulfur. To emphasize the importance of XUV + X-ray radiation, we provide comparisons with
the case when only X-ray radiation is considered. Although unobservable, we believe an
observational signature of XUV radiation has already been detected by M03.  In their {\it FUSE}
study of LMC stars, they explicitly state that ``...\CV\ is the dominant species for O stars''.  This
can only mean that there must be excess XUV emission since the \CIV\ photoionization edge lies
within the XUV energy range (see Fig. 2). 

We first discuss the expected contribution of XUV radiation to the radiative cooling curve (e.g.,
Cox \& Tucker 1969), $\Lambda(T) = P(T)/(n_e~n_H$) ( erg cm$^{3}$ s$^{-1}$) where
$P(T)$ is the power per unit volume and $n_e$ and $n_H$ are the electron and hydrogen number
densities respectively. Although cooling curves have undergone alterations over the years (e.g.,
atomic data updates and added emission lines), the basic shape of the cooling curve has remained
intact as shown in Figure 1. This shows $\Lambda(T)$ obtained from Raymond \& Smith (1977;
RS), Mewe \etal\ (1985; MEKAL), and Smith \etal\ (2001; APED) data. It also shows that the
XUV contribution to $\Lambda(T)$ is clearly important for temperatures between 0.5 and 2.0
MK. 

Since the photoionization edge for \PV\ $\rightarrow$ \PVI\ is at 65.03 eV, and that for \SV\
$\rightarrow$ \SVI\ is nearby at 72.68 eV, it seems plausible that their ionization balances should
be quite similar. This similarity led to the F06 argument that \SV\ can be used as a surrogate for
\PV. However, as shown in Figure 2, just beyond the \PV\ photoionization edge there exists a
large collection of intense XUV emission lines (their specific contribution is shown in Fig. 1) that
are located {\it just below} the \SV\ ionization energy. Whereas, at energies higher than this edge,
the XUV emission at this temperature is essentially devoid of emission lines, i.e., all lines between
the \SV\ edge and 124 eV are at least a factor of 100 times smaller than the strongest line just
below the \SV\ edge. Consequently, this unique energy distribution of XUV lines relative to the
energy locations of these photoionization edges indicates that the XUV emission should produce
different effects on the phosphorous and sulfur ionization equilibria. 
This is illustrated in Figure 3 which shows that the \PV\ photoionization 
rate is $\approx$ 10 times larger than the \SV\ rate at the temperature 
of maximum XUV emission (see Fig. 1). Figure 3 includes the \CIV\ rate for
comparison, and also shows the expected photoionization rates using only the  
X-ray energy band ($\ge$ 124 eV). As evident from the displacement of these  
curves, the neglect of XUV radiation leads to underestimates of these 
rates which are substantial at temperatures $<$ 2 MK. This implies that alone, the 
radiation from the X-ray energy band is not expected to have a significant impact
on the ionization equilibrium of phosphorous, as was recently demonstrated by 
Krticka \& Kubat (2009) in their study of the Auger effect on the fractional
abundance of \PV.

Our graphical arguments illustrate that the XUV line emission for temperatures between 0.5 to 2
MK is expected to have a major impact on the ionization structure of phosphorous but a relatively
minor effect on sulfur. This implies that the sulfur surrogate argument needs to be re-examined
(see \S 3). In fact, since the \CIV\ photoionization edge is almost identical to the \PV\ edge (see
Fig. 2), \CIV\ is a more appropriate surrogate for \PV. The main difference is that the \CIV\
photoionization rate is larger than the \PV\ rate (see Fig. 3) due to the differences in their cross
sections.

\section{Ionization Equilibrium Calculations and Required XUV Radiation}
The stellar wind ionization equilibria of phosphorous and sulfur are calculated in a straightforward
way to determine the level of XUV+X-ray radiation required to affect the fractional ionization
abundances, i.e., \QPV\ and \QSV. We consider a stellar effective temperature (\Teff) range from
27500 to 45000 K which covers the O and early B spectral range where \PV\ has been used to
study \Mdot. The ionization equilibrium is determined by adopting an ionization/recombination
rate balance approach similar to the one used in FASTWIND (Puls et al. 2005). The main
differences are: 1) we use the photoionization cross sections of Verner \& Yakovlev (1995); 2) a
wind diffuse field as prescribed by Drew (1989), and; 3) we assume a radial power law dependent
wind temperature ($T_W$) which is adjusted to produce a phosphorous wind ionization structure
similar to that of Puls \etal\ (2008) (i.e., no XUV+X-ray radiation) for all \Teff\ considered. This
adjustment leads to a $T_W / \Teff = 1.15 (\Rstar/r)^{0.5}$ with a minimum value of $0.6\Teff$.
We calculate \QPV\ and \QSV\ for wind conditions at a location where the wind velocity is $1 /
2$ the terminal velocity, \vinf.  For each \Teff, a consistent set of stellar parameters (\logG, \Lbol,
\Rstar) is found by using the fitting-formulae given by Martins \etal\ (2005). We use the electron
scattering Eddington factor as described by Lamers (1981) to determine the effective escape
velocity, \Vesc. For the wind parameters, we use $\vinf = 2.6 ~ \Vesc$ and \Mdot\ predicted by
the Vink \etal\ (2000) formula. 
The radially dependent wind density is determined from the mass conservation 
equation using a $\beta$-velocity law $V(r) = \vinf (1 - R_*/r)^{\beta}$ 
assuming a $\beta = 0.8$ (Pauldrach \etal\ 1986; M\"{u}ller\ \& Vink 2008). 
The wind electron density is derived assuming hydrogen and helium are fully ionized. 

The energy dependent mean intensity (erg cm$^{-2}$ s$^{-1}$ eV$^{-1}$ str$^{-1}$) used to
calculate the photoionization rates has two dominant contributions: 1) the photospheric radiation
field for a given \Teff\ and appropriate \logG\ (using the TLUSTY grid of models from
OSTAR2002, Lanz \& Hubeny 2003), along with the standard geometric dilution factor, and; 2)
the XUV+X-ray radiation field, \JX. This \JX\ is specified by two parameters, the hot plasma
temperature, \TX, and the column emission measure, \DEMC\ (cm$^{-5}$) such that $\JX (E,
\TX) = \DEMC ~ \epsilon (E,\TX)$ where $\epsilon$ is the energy-dependent emissivity (erg
cm$^3$ s$^{-1}$ eV$^{-1}$ str$^{-1}$) taken from the APED data. 
The basic assumption is that \JX\ represents an ``effective'' XUV+X-ray 
mean intensity at each given radial location, i.e., the wind contains a finite but 
small level of XUV+X-ray radiation distributed throughout the wind which 
seems to be supported by observations as discussed in \S 2.
All calculations presented in this section use a $\TX = 1$ MK so we can 
examine the maximal effects of XUV radiation on \QPV\ and \QSV.
The total input mean intensity is determined for an energy grid  
from 8 eV to 2 keV.

Since we are concerned with studying ionization effects for all
luminosity classes over a large range in \Teff\ for which there is \PV\ data, 
it is advantageous to define a new parameter that is dependent on both
X-ray and stellar parameters. 
By defining \FX\ as the energy integral 
of $4 \pi \JX(E,\TX)$ above the \HeII\ edge (erg cm$^{-2}$ s$^{-1}$), then 
our fundamental adjustable parameter used in this study is defined as \FXFS\ 
where $F_*$ is the total stellar photospheric flux ($\Lbol / 4 \pi \Rstar^2$). 
Hence, for a given $F_*$, \FXFS, and \TX, \DEMC\ can be 
extracted directly from 
\begin{equation}
\DEMC = { \left ( \frac {\FX} {F_*} \right ) } ~ \frac {F_*} {\Lambda (\TX)},
\end{equation}
where $\Lambda (\TX)$ (erg cm$^3$ s$^{-1}$)  is the total energy integral 
of $4 \pi \epsilon (E, \TX)$ above the \HeII\ edge. Note that \FXFS\ is not 
the same as the well known observed X-ray to bolometric luminosity ratio, 
\LXLB, because \LX\ in this ratio is an ``observed'' quantity, i.e., a measure 
of only those X-rays capable of escaping the stellar wind, and our \FX\ is
defined as an intrinsic total mean intensity (in flux units) where the majority
of this emission (i.e., XUV) resides in an observational window that is 
inaccessible due to wind and ISM attenuation.

%

The predicted \QPV\ dependence on \FXFS\ for supergiants, giants, and main sequence stars is
shown in Figure 4. Also shown in this Figure are the data points from F06 that correspond to the
required \QPV\ values if all stars have their traditional \Mdot\ which we will use to determine the
constraints on \FXFS. This deficit in \QPV\ relative to unity led F06 to conclude that \Mdot\
needs to be reduced. 
Figure 4 shows a strong dependence of \QPV\ on \FXFS, and indicates that a
range in \FXFS\ between $(0.3 ~ - ~ 10) \times 10^{-7}$ can explain 
the observed \QPV\ for all luminosity classes.  
For example, from Figure 4, the observed \QPV\ of the four supergiants at 
$\Teff = 35000$ indicate a $\FXFS \approx 2.5 \times 10^{-7}$.
This implies a XUV+X-ray flux $ \approx 2 \times 10^{7}$ erg cm$^{-2}$ s$^{-1}$ 
and a \DEMC\ $\approx 2.5 \times 10^{29}$ cm$^{-5}$ (using Eq. 1).
From X-ray analyses of OB stars we cannot directly determine \DEMC\ since 
only the volume emission measure $EM_V$ (cm$^{-3}$) can be 
deduced from observations. If we assume that the XUV+X-ray radiation arises 
from a spherically shell at the assumed radius ($r = 1.7\Rstar$ where $\Rstar =
20.5\Rsun$), then the {\it ``intrinsic''} $EM_V$ is $\approx 1.8 \times 10^{55}$
(using $4 \pi r^2 ~ \DEMC$) which is comparable to the
lowest energy line {\it ``observed''} $EM_V$ ($\approx 5 \times 10^{55}$ 
for \NVII) derived from {\it Chandra HETGS} observations 
(e.g., Wojdowski \& Schulz 2005). Since the {\it intrinsic} $EM_V$
is expected to be $>$ the {\it observed} $EM_V$ (optical depth effects), 
the XUV+X-ray flux required to reduce \QPV\ is well within the 
observational limits.


Now we examine the effects of XUV+X-ray radiation on \SV\ by considering the ratio 
$\QPV / \QSV$. As shown in Figure 5, for the case when $\FX= 0$, this ratio is 
$\approx 1$ for a large range in \Teff\ which supports the F06 \SV\ surrogate argument.
However, as \FXFS\ increases, $\QPV / \QSV$ decreases which means that \QSV\ 
is significantly less sensitive to the XUV+X-ray radiation as compared to the 
dependence of \QPV. This is a direct consequence of the difference in the \PV\ 
and \SV\ photoionization rates shown in Figure 3 (see \S 2), and invalidates the 
sulfur surrogate argument. In general, for all luminosity classes,  
$\QPV / \QSV$ reaches a minimum value of $\approx 0.14$ over most of 
the O-star spectral range. The required \FXFS\ to produce this minimum 
value is dependent on luminosity class as shown in Figure 5. These results imply 
that the \Mdot\ derived from {\it FUSE} observations could be underestimated 
by almost an order of magnitude.


\section{Discussion}
We have demonstrated that the presence of XUV radiation embedded throughout a stellar wind in
sufficient amounts (well within the observational constraints) can lead to a significant depletion of
\QPV\ by using only outer shell photoionization processes. Therefore, the
discrepancy between \PV\ derived \Mdot\ with those obtained from density-squared diagnostics is
far less severe than suggested by F06. Hence, these stars can have \Mdot\ that are again back
within a factor of 2 range of the traditional \Mdot\ (consistent with clumped wind predictions). 
This also means that the XODUR does not require an alternative explanation. 

Gudel \& Naze (2009) mentioned that there may be a problem with XODUR 
since the wind opacities used by Waldron \& Cassinelli (2007) were determined 
from a more highly ionized wind. 
Although it is true that the opacity at {\it low energies} can be
sensitive to the assumed wind ionization structure, the key point is that the XODUR 
is based entirely on emission lines at energies $\geq$ 0.56 keV where the 
opacity is almost identical to the ``cold'' ISM opacity above this energy
(e.g., see Fig. 2 of Waldron \etal\ 1998), and the ISM opacity
represents the unsurpassable upper limit to any wind opacity. 
Therefore, regardless of the wind ionization state, the XODUR is certainly 
valid for those radii derived from the He-like $fir$ lines of \NeIX, \MgXI, 
\SiXIII, and \SXV, and radii derived from \OVII\ may be marginally 
dependent on the wind ionization state.

We have shown that the inclusion of XUV radiation can resolve the \Mdot\ discrepancy, and there
is observational support that this XUV radiation must be distributed throughout these winds
based on the conclusion of M03 regarding \CV. But, what is the source of this emission? Since
these winds are clumped, this emission is likely to originate from bow shocks forming around
clumps. The Cassinelli \etal\ (2008) wind bow shock model predicts that the temperature
dependence of the emission measure scales as $(T/T_{max})^{-4/3}$, where $T_{max}$ is the
temperature at the apex of the bow shock. Hence, there should be a considerable amount of XUV
radiation, even for rather strong shocks, located at any radius. Although this XUV radiation is
unobservable since essentially all of this radiation will be absorbed by the stellar wind and ISM,
the strength of this emission can be determined indirectly from analyses of lower energy spectral
bands as demonstrated in this paper.

Future studies now need to focus on alternate explanations as to why the X-ray 
lines are symmetric and nearly un-shifted. Several possibilities have been proposed: 
wind porosity effects (Oskinova \etal\ 2004, 2006); asymmetric mass outflows 
(Mullan \& Waldron 2006), and; bow shocks forming around wind clumps 
(Cassinelli \etal\ 2008) and stellar ejected plasmoids (Waldron \& Cassinelli 2009). 
In addition, Oskinova \etal\ (2007) claim that the X-ray line symmetry and \PV\ problems 
can both be resolved, with no need to reduce \Mdot, using their planar shock model, 
porosity effects, and macro-clumping. However, we argue that bow shocks will 
undoubtedly form around these macro-clumps and the impact of the resultant 
excess XUV radiation on \QPV\ needs to be examined. Regardless as to which 
explanation is applicable, our work has re-emphasized the importance of 
including the full XUV+X-ray energy spectrum when exploring the effects 
of radiation on stellar wind ionization structures. 


\acknowledgments
This work was supported by NASA ATFP award NNH09CF39C.

\clearpage
\begin{figure}
\includegraphics[height=17cm,angle=-90]{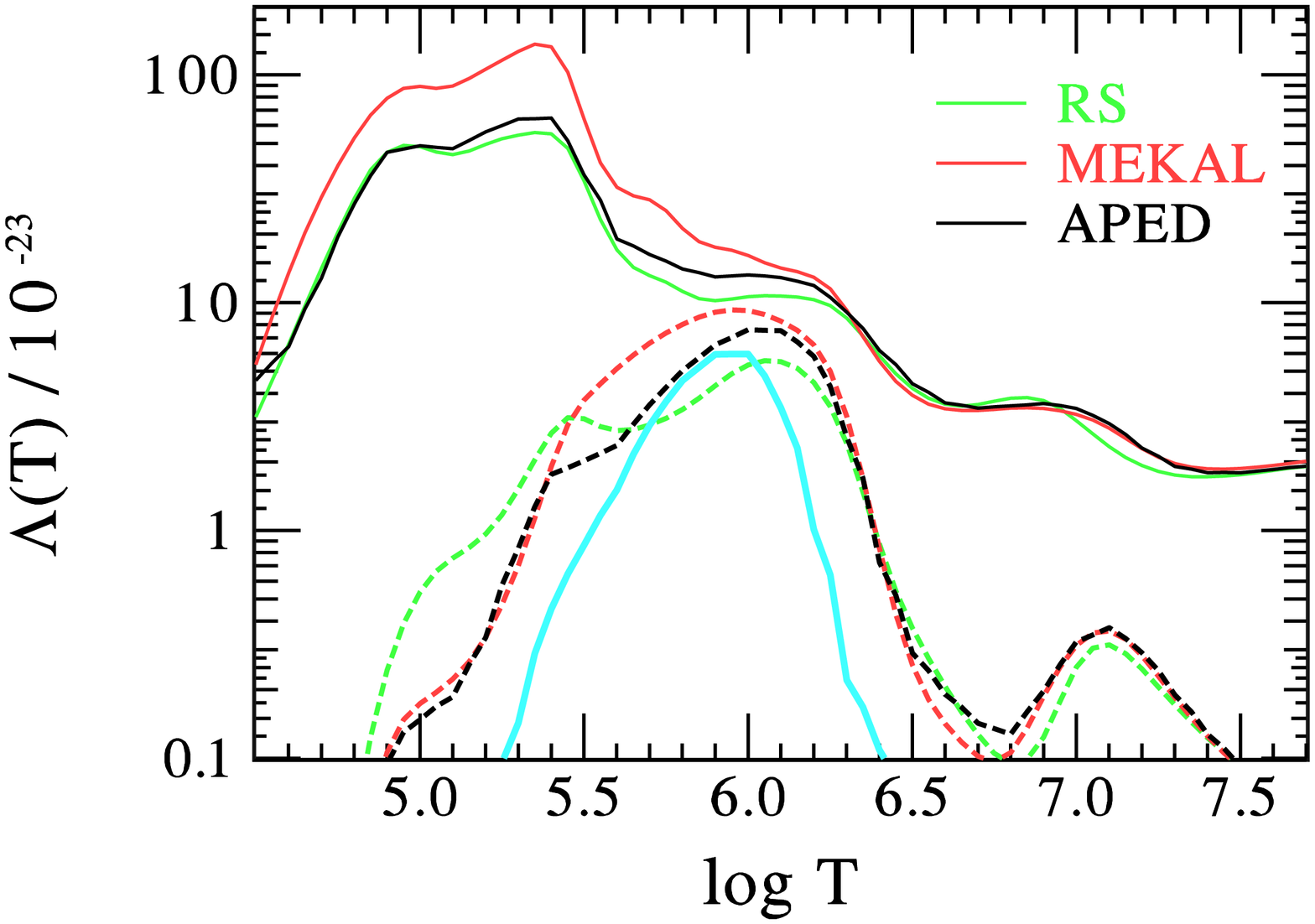}
\caption{Comparison of radiative cooling curves determined from the RS, MEKAL, and APED
emissivites.  The dashed-line curves represent the contribution of the XUV radiation (54.4 - 124
eV) to the total cooling curve. The solid blue line represents the contribution from the important
collection of XUV emission lines between 65 and 73 eV (See \S 2 and Fig. 2).
\label{fig:EMISFULL}}
\end{figure}

\clearpage
\begin{figure}
\includegraphics[height=17cm,angle=-90]{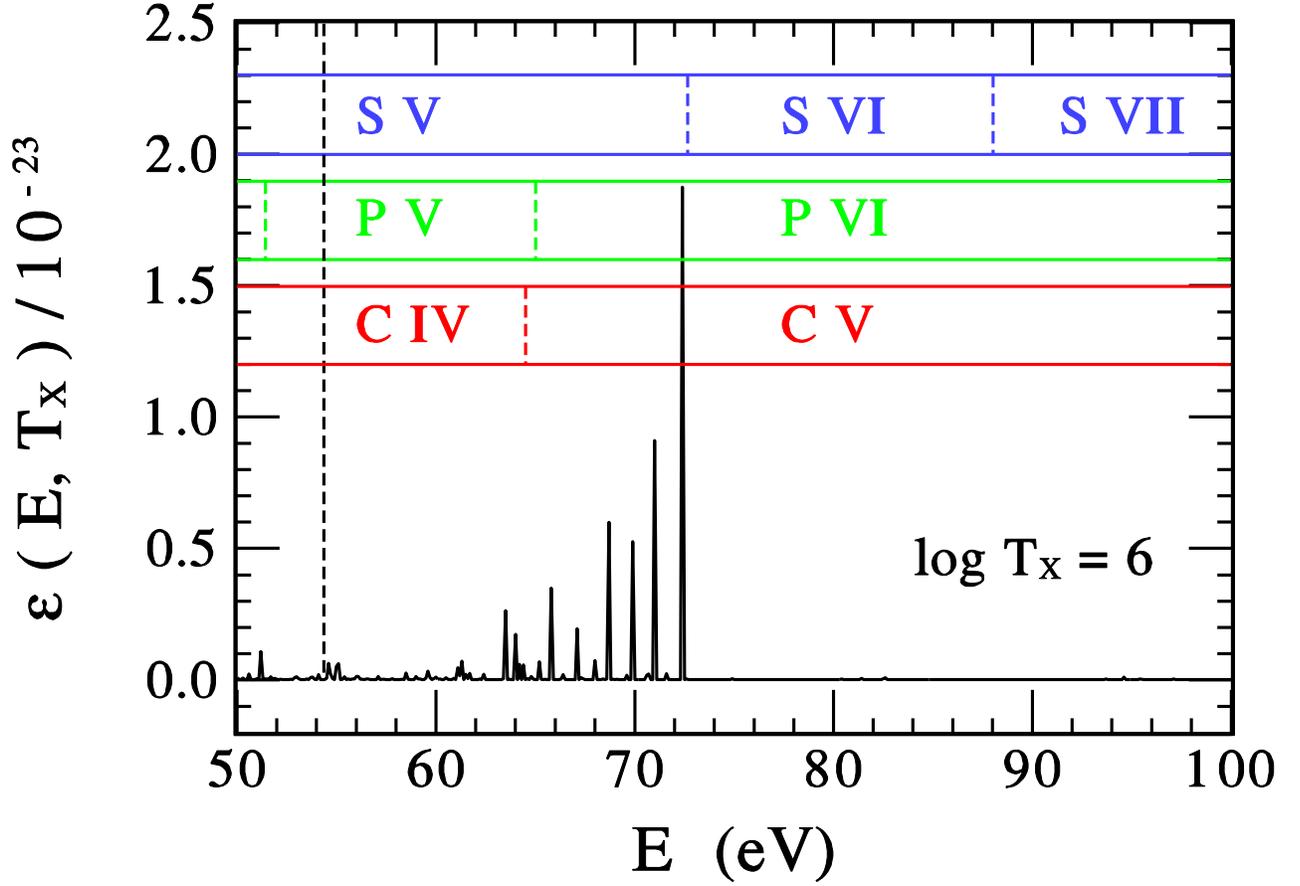}
\caption{Comparison of the energy dependent X-ray emissivity ($\TX = 1$ MK) with the
photoionization energy ranges of the relevant ions of carbon, phosphorous, and sulfur. This shows
the strong collection of XUV emission lines just below the \SV\ threshold energy at 72.68 eV, and
above this energy, the XUV radiation is essentially devoid of emission lines (see \S 2). The four
dominant XUV lines are Fe IX (72.47 eV), Fe X (71.04, 69.95 eV), and Fe XI (68.72 eV). The
vertical dashed line is the \HeII\ edge (54.4 eV).
\label{fig:IONEDGE}}
\end{figure}

\clearpage
\begin{figure}
\includegraphics[height=17cm,angle=-90]{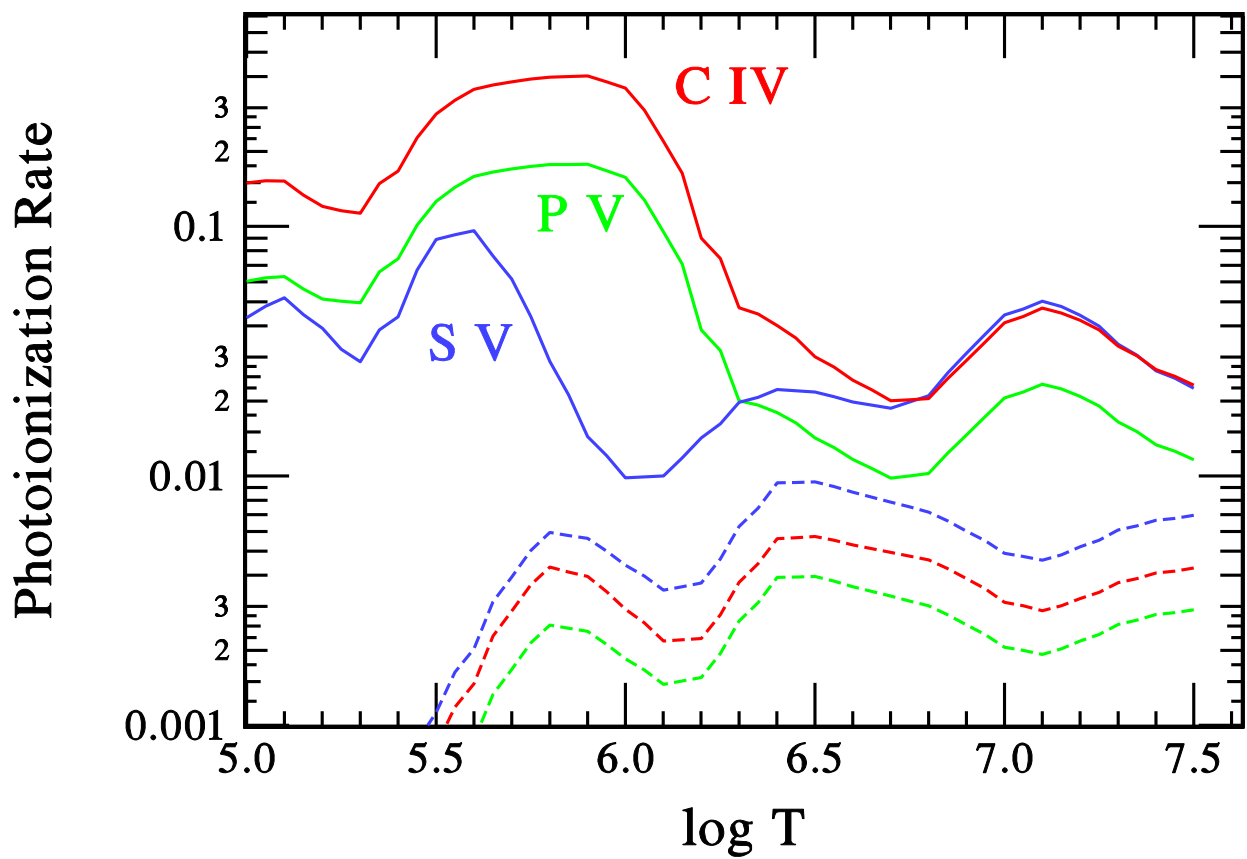}
\caption{
The XUV+X-ray photoionization rates for \CIV, \PV, and \SV\ as a 
function of temperature assuming that $4\pi$ times  the total 
energy integrated XUV+X-ray mean intensity above 54.4 eV 
is fixed at $10^8$ erg cm$^{-2}$ s$^{-1}$ for each temperature.  
The dashed-line curves show these rates as determined by using 
only the X-ray energy band ($\ge$ 124 eV).
\label{fig:XP_RATE}}
\end{figure}

\clearpage
\begin{figure}
\includegraphics[height=17cm,angle=-180]{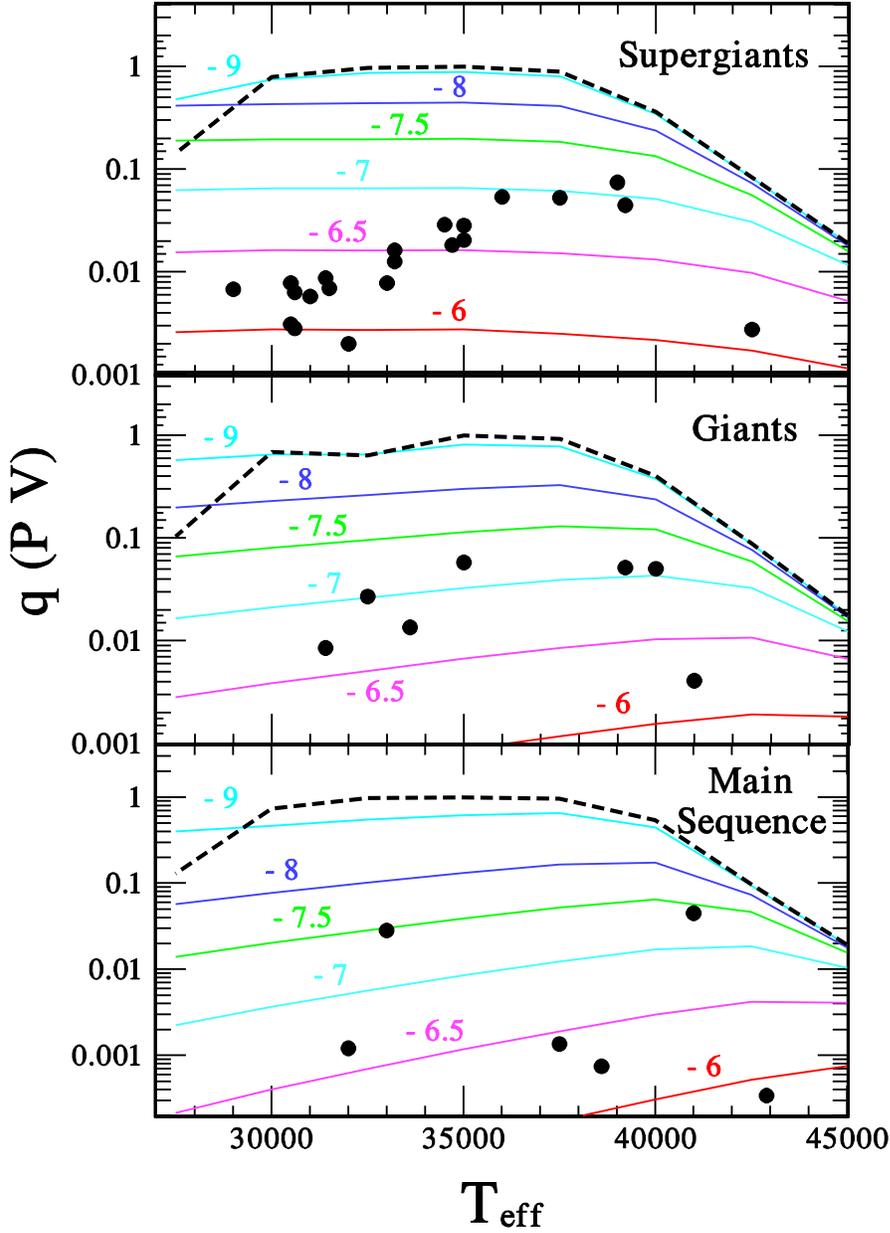}
\caption{The effects of XUV radiation on \QPV\ as a function of \Teff\ for different values of
log (\FXFS) for the three luminosity classes. The black dashed-line represents the $\FX = 0$ case.
The filled circles are the observationally determined \QPV\ assuming traditional \Mdot\ from F06
(see \S 3). 
\label{fig:PVXFLUX}}
\end{figure}

\clearpage
\begin{figure}
\includegraphics[height=17cm,angle=-180]{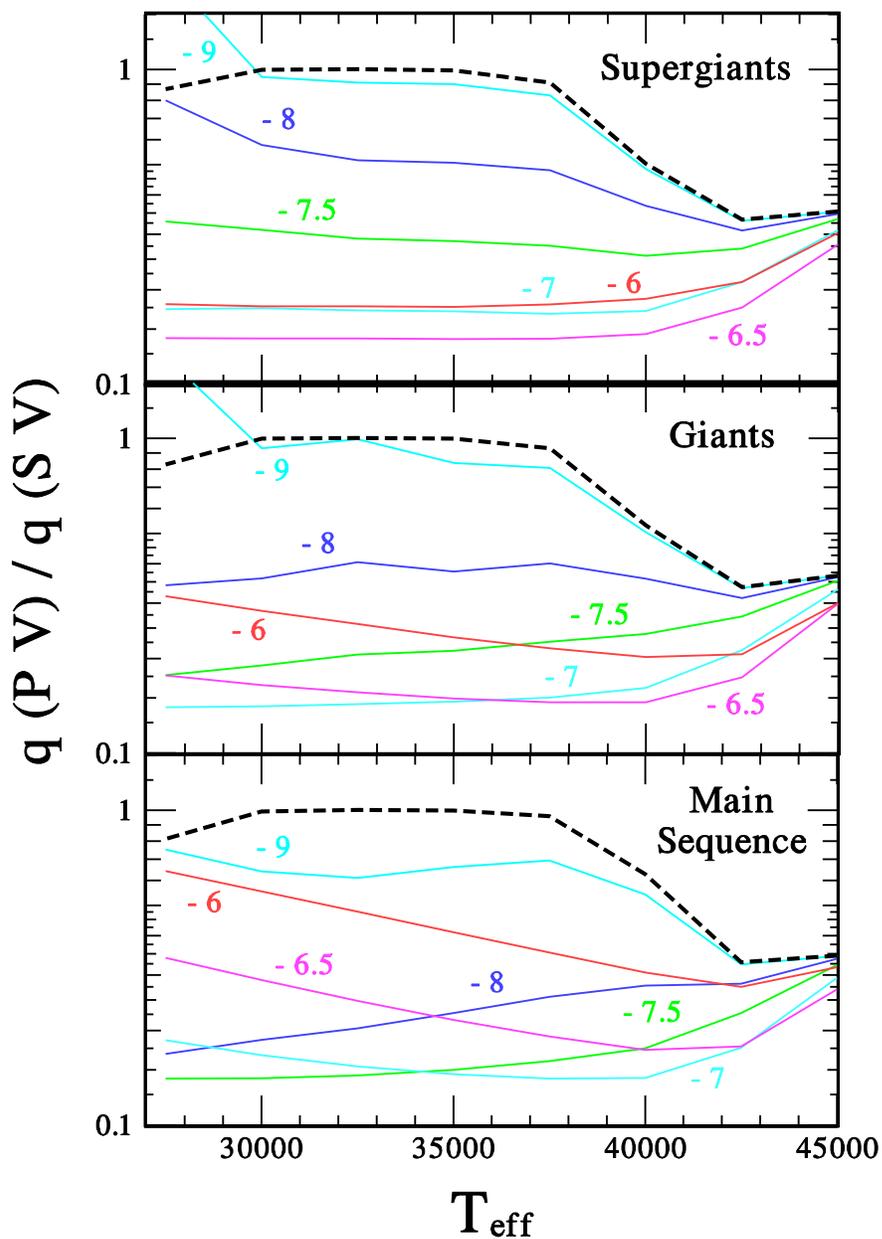}
\caption{The ratio of $\QPV / \QSV$ as a function of \Teff\ for different values of log (\FXFS)
for the three luminosity classes. The black dashed-line represents the $\FX = 0$ case and this
ratio is $\approx 1$ for $\Teff < 39000$ K. See discussion in Section 3.
\label{fig:SVPVFX}}
\end{figure}

\end{document}